\documentclass[aps,pra,twocolumn, showpacs,
floatfix]{revtex4-2}

\usepackage[utf8]{inputenc} 
\usepackage{amssymb,amsmath}
\usepackage{graphicx}
\usepackage{braket}
\usepackage{siunitx}
\usepackage{xcolor}
\usepackage[normalem]{ulem}
\usepackage{bm}

\def\Rb87{^{87}\mathrm{Rb}}                     

\def\ex{\mathbf{e}_x}  
  
\def\ez{\mathbf{e}_z}  
\def\kb{k_{\rm B}}
\def\kr{k_0}
\def\Er{E_0}

\newcommand{\M}{\ensuremath{\hat{M}({\bf k}_\perp)}} 

\usepackage{verbatim}

\begin{document}

\title{Weak-Measurement-Induced Heating in Bose-Einstein Condensates}

\author{Emine~Altunta\c{s}}
\email{altuntas@umd.edu}
\affiliation{Joint Quantum Institute, National Institute of Standards and Technology, and University of Maryland, Gaithersburg, Maryland, 20899, USA}
\author{I.~B.~Spielman}
\email{ian.spielman@nist.gov}
\affiliation{Joint Quantum Institute, National Institute of Standards and Technology, and University of Maryland, Gaithersburg, Maryland, 20899, USA}
\homepage{http://ultracold.jqi.umd.edu}
\date{\today}

\begin{abstract}
Ultracold atoms are an ideal platform for understanding system-reservoir dynamics of many-body systems. 
Here, we study quantum back-action in atomic Bose-Einstein condensates, weakly interacting with a far-from resonant, i.e., dispersively interacting, probe laser beam.
The light scattered by the atoms can be considered as a part of quantum measurement process whereby the change in the system state derives from measurement back-action.
We experimentally quantify the resulting back-action in terms of the deposited energy.
We model the interaction of the system and environment with a generalized measurement process, leading to a Markovian reservoir.
Further, we identify two systematic sources of heating and loss: a stray optical lattice and probe-induced light assisted collisions (an intrinsic atomic process).
The observed heating and loss rates are larger for blue detuning than for red detuning, where they are oscillatory functions of detuning with increased loss at molecular resonances and reduced loss between molecular resonances.
\end{abstract}

\maketitle


In recent years, there have been rapid breakthroughs in quantum technologies that offer new opportunities for advancing the understanding of basic quantum phenomena; realizing novel strongly correlated systems~\cite{Semeghini2021}; and enhancing applications in quantum communication, computation, and sensing~\cite{Asenbaum2020}.
Cutting edge applications require high fidelity quantum measurement and control.
Qubit based quantum error correction~\cite{Terhal2015} is a prominent example where both of these elements are indispensable for first measuring the state of ancilla qubits and then applying the requisite feedback~\cite{Negnevitsky2018, Livingston2022}.
Quantum metrology provides a second example, where the combination of measurement and feedback enables the generation of squeezed states with metrologically useful entanglement~\cite{Cox2016, Murray2019} and deterministic entanglement in superconducting qubits~\cite{DiCarlo_2013}.
Quantum feedback control of ultracold atoms is a new direction that relies on this toolbox as a means to engineer new dynamical steady states that cannot be achieved in a closed equilibrium system~\cite{Ivanova2020, Kopylov2015, Kroeger_2020}.
All together these designate back-action limited measurements as essential for fully cultivating this platform’s ultimate potential. 

Here we study quantum back-action in atomic Bose-Einstein condensates (BECs), weakly measured by a far-detuned probe laser beam.
All quantum measurements, no matter how weak, partially collapse the system's wave function into the state indicated by the measurement outcome.
Thus the act of measurement imparts energy to an equilibrium system, making heating a diagnostic of quantum back-action.
In this paper we experimentally characterize measurement-induced heating, and in addition report two key parasitic effects resulting from the measurement process: light-induced collisions and a stray optical lattice formed from the probe beam itself. 

We also outline a quantum trajectory based measurement model, and focus on the information extracted by light-scattering as a measurement process.
This model describes experiments in which the scattered light is measured by the environment, and the associated back-action---here heating---on the system is experimentally observed.
In a companion paper, back-action is further characterized using Ramsey interferometry, probing measurement-induced decoherence~\cite{Altuntas2022_RIL}.

Weak measurements---sometimes termed partial or non-destructive---enable dynamically monitoring of a single quantum system. 
Established cold-atom applications include the observation of real-time vortex dynamics~\cite{Freilich1182, Ferrari_PRL2015}, spinor dynamics~\cite{HigbieSadler2005}, and the formation of ferromagnetic order in spinor BECs~\cite{Sadler2006}.
In these studies it was sufficient that the disturbance from each measurement did not appreciably influence the relevant dynamics, as they focused on mean-field dynamics in large atom number BECs.

To date the ultimate quantum back-action limit of such weak measurement techniques have not been considered for quantum gas experiments except in optical cavities~\cite{Murch2008a}.
Back-action limited measurements alone enabled the production of squeezed spin states in optical cavities, with the overall collective spin conditioned on the measurement outcome~\cite{Vuletic2010,Thompson2014,Hosten2016}.
Similarly recent work on partially measured qubit arrays indicate the existence of entanglement transitions that can only be identified given knowledge of measurements~\cite{GullansPRX_20,Crystal_22}.
Going beyond these measurement applications, closed loop quantum control is also reliant on back-action limited measurements.
Even at their most simple, measurement-based feedback schemes provide a potential alternative cooling method for ultracold atoms~\cite{Mitchell_2013, Hush_2013, Schemmer2017}. 
A single-stage feedback scheme reducing number fluctuations was demonstrated in Ref.~\cite{Gajdacz2016}; however, it was applied to a thermal gas where quantum-limited measurements were not required.
Further, generalized schemes are predicted to drive quantum gases to new many body phases~\cite{Mazzucchi2016,Mazzucchi2016a,Young2021a,Lloyd2000}.
From this perspective, characterizing the limits to quantum measurement is a first step towards implementing feedback control and measurement-driven Hamiltonian engineering in many-body systems.

This paper begins by reviewing our measurement-based theoretical framework in Sec.~\ref{sec:MeasurementModel}.
We elaborate on the details of bolometry for determining measurement-induced heating in Sec.~\ref{sec:Bolometry}.
Next, in Sec.~\ref{sec:ExpOverview}, we continue with a brief description of our experimental setup and methods.
Section~\ref{sec:Photoassiciation} discusses the effects of light assisted collisions on loss [App.~\ref{app:PCI_signal} expands on this discussion by presenting the phase-contrast imaging (PCI) data corresponding to these measurements.].
In Sec.~\ref{sec:MI_heat_results}, we study the added energy by measurement and discuss the ensuing systematic effects.
Section~\ref{sec:Conc} concludes with a discussion of experimental and theoretical implications, and describes potential future directions for research.


\section{Model} \label{sec:MeasurementModel}

\begin{figure}[htb!]
\includegraphics{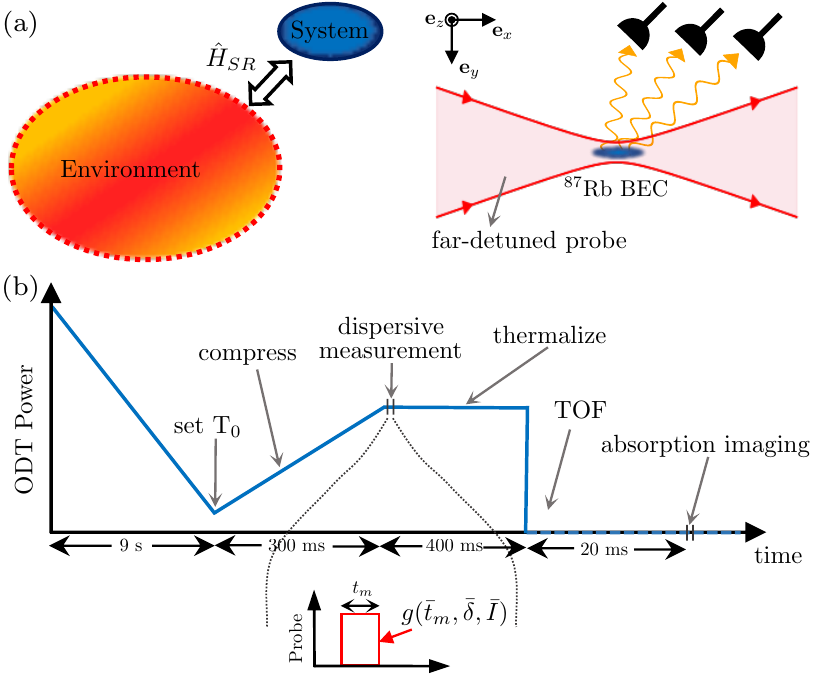}
\caption{Model schematic and experimental concept.
(a) System-reservoir interaction. 
Left: General concept. 
The system (BEC) is coupled to the reservoir via the interaction Hamiltonian $\hat H_{\rm SR}$. 
Right: Experimental concept. 
The BEC (blue) is illuminated with far-detuned laser light (red) and scatters light (wiggly orange lines) into both occupied and empty reservoir modes.
The reservoir modes are then projectively measured by the environment, modeled by an array of photo-detectors.
(b) Time sequence employed to determine the measurement-induced heating. 
The ODT power (vertical axis) was initially ramped down to establish $T_0$ yielding either a BEC or a thermal gas. 
Subsequently the trap power was ramped up in 300~ms, and immediately following this compression stage, the far-detuned probe beam (with $1/e^2$ minimum waist $\approx 700~\mu{\rm m}$ aligned to the BEC) illuminated the atomic cloud for a time $t_{\rm m}$ realizing system-reservoir coupling.
Following a 400~ms thermalization period, the ODT was turned off, initiating TOF.
After 20~ms of TOF expansion, the momentum-distribution of the atomic cloud was detected via standard absorption imaging. 
}
\label{Fig1:MeasurementModel}
\end{figure}

We outline our theoretical approach presented in Ref.~\cite{Altuntas2022_RIL} focusing on a weakly interacting atomic BEC (the system) dispersively coupled to optical electric fields $\hat{\bf E}({\bf x},t)$ (the reservoir) as illustrated in Fig.~\ref{Fig1:MeasurementModel}(a).
In this model, the interaction picture system-reservoir Hamiltonian
\begin{align}
\hat H_{\rm SR}(t) &=\! \int\! \frac{d^3 {\bf x}}{\hbar\Delta} \hat n_{\rm g}({\bf x})\otimes
[\hat{\bf E}({\bf x},t)\!\cdot\!{\bf d}_{\rm ge}][{\bf d}_{\rm ge}^{\ast}\!\cdot\!\hat{\bf E}^\dagger({\bf x},t)],
\label{Eqn:acStark_Hamiltonian}
\end{align}
describes the interaction of light with two-level atoms, giving the ac Stark shift to the atoms and a dispersive phase shift to the light. 
Here $\hat n_{\rm g}({\bf x}) = \hat{b}_{\rm g}^{\dagger}({\bf x}) \hat{b}_{\rm g}({\bf x})$ is the atomic density operator expressed in terms of the bosonic field operators $\hat{b}_{\rm g}({\bf x})$ for ground state atoms at position ${\bf x}$; ${\bf d}_{\rm ge}$ is the dipole matrix element for transitions between ground and excited state atoms with energy difference $\hbar \omega_{\rm ge}$; and lastly, $\Delta = \omega_0 - \omega_{\rm ge}$ is the detuning from atomic resonance of a probe laser with frequency $\omega_0$. 

In the limit $|\Delta|\ll\omega_{\rm ge}$, we express the optical electric field operator 
\begin{align*}
\hat {\bf E}({\bf x},t) &= i \sqrt{\frac{\hbar \omega_{\rm ge}}{2 \epsilon_0}}  \sum_{\sigma} \int\!\frac{d^3{\bf k}}{(2\pi)^3} \hat a_\sigma({\bf k}) \boldsymbol{\epsilon}_{\sigma}({\bf k}) e^{i ({\bf k}\cdot {\bf x}-c|{\bf k}| t )},
\end{align*}
in terms of field operators $\hat{a}_{\sigma}({\bf k})$ describing states with wave vector ${\bf k}$ and polarization $\sigma$.
Here, $\epsilon_0$ is the electric constant; $c$ is the speed of light; and $\epsilon_{\sigma}({\bf k})$ are a pair orthogonal polarization vectors transverse to ${\bf k}$, labeled by $\sigma = \pm$. 
Here, each outgoing mode is in a specific polarization state ${\epsilon}({\bf k}_\perp)$ rendering the polarization subscript redundant.

We assume a probe laser of wavelength $\lambda$ occupies a single optical mode $({\bf k}_0, \sigma_0)$ with $\kr \equiv |{\bf k}_0| = 2\pi/\lambda$.
This ansatz enables us to make the replacement $\hat a_\sigma({\bf k}) \rightarrow \delta({\bf k}-{\bf k}_0) \delta_{\sigma,\sigma_0} \alpha_0 + \hat a_\sigma({\bf k})$, which describes a coherent driving field with amplitude $\alpha_0 \gg 1$.
In this expression the modes $\hat a_\sigma({\bf k})$ are initially empty.
With this replacement, we expand Eq.~\eqref{Eqn:acStark_Hamiltonian} in decreasing powers of the large parameter $\alpha_0$.
The leading term corresponds to the ac Stark shift, and the next term 
\begin{align*}
\hat H_{\rm eff} &= \frac{\hbar P_{\rm e}^{1/2}}{(ct_{\rm m})^{1/2}}\! \oint_{k_0}\! \frac{d^2 {\bf k}_\perp}{(2\pi)^2}
g^*({\bf k}_\perp) 
\hat{n}_{\mathcal{F}}({\bf k}_\perp\!-{\bf k}_0)\hat a^\dagger({\bf k}_\perp) + \rm{H.c},
\end{align*}
describes scattering from the probe field into optical modes by spatial structure in the atomic density, where $P_{\rm e} = |\alpha_0 g_{\sigma_0}({\bf k}_0)|^2 / \Delta^2 $ is the excited state occupation probability.
In the far-detuned limit, the outgoing wavenumber is fixed at $k_0$ leading to the surface integral over the sphere of radius $k_0$.
The coupling constant
\begin{align}
g({\bf k}_\perp) & \equiv -i \left(\frac{\omega_{\rm ge}}{2\hbar \epsilon_0} \right)^{1/2} \left[{\bf d}_{\rm ge} \cdot \boldsymbol{\epsilon}({\bf k}_\perp)\right],
\label{Eqn:Couple_Strength}
\end{align} 
quantifies the coupling strength between the incident monochromatic light and an outgoing mode of wave vector ${\bf k}_\perp$. 

The Fourier components of the density distribution
\begin{align*}
\hat{n}_{\mathcal{F}}({\bf k}_\perp-{\bf k}_0) = \int \frac{d^3 {\bf k}}{(2\pi)^3} \hat b^\dagger[{\bf k} - ({\bf k}_\perp-{\bf k}_0)] \hat b({\bf k}).
\end{align*}
indicate that the back-action on the atomic system from a photon recoiling in direction ${\bf k}_\perp$ places each atom into a coherent superposition with amplitude recoiling in the opposite, momentum conserving, direction.

We model the larger environment as an observer that measures the outgoing light in the far-field with an ideal photo detection process [Fig.~\ref{Fig1:MeasurementModel}(a)], that is, a strong measurement of the photon density $\hat{a}^\dagger({\bf k}_\perp) \hat{a}({\bf k}_\perp)$. 
This measurement process begins with the combined system-reservoir state $\ket{0_{\rm R}}\otimes\ket{\Psi_{\rm S}}$, describing a reservoir with no photons (other than those in the probe) and with the system in an arbitrary state.
This state briefly evolves for a time $t_{\rm m}$ via the time evolution operator $\hat{U}_{\rm SR}(t_{\rm m}) = \mathcal{T} \exp\left[-i \int_{-t_{\rm m}/2}^{t_{\rm m}/2} \hat{H}_{\rm {eff}}(t) d t/\hbar\right]$
at which time a photon may be detected in momentum state ${\bf k}_\perp$, and the atomic system correspondingly recoils. 
Altogether, this two step protocol constitutes a generalized measurement of the system realized by a projective measurement of the reservoir.
In this schema, the conditional post-measurement system wave function $\ket{\Psi_{\rm S}^\prime}$ is described by the Kraus operator $\M = \bra{{\bf k}_\perp} \hat{U}_{SR}(t_{\rm m})\ket{0_{\rm R}}$ yielding $\ket{\Psi_{\rm S}^\prime } = \M \ket{\Psi_{\rm S}}$.

Intuitively, there are two scattering mechanisms: either light scatters from the ``actual'' density distribution $\langle \hat n_{\rm g}({\bf x}) \rangle$, or from quantum fluctuations around that mean.
In the case of a BEC, these lead to stimulated and spontaneous scattering, respectively.
In the context of heating and back-action, stimulated scattering imparts a collective response, that results in ``lensing'' of the scattered light~\cite{Andrews1996}.
In our experiments this effect is negligible and heating is dominated by spontaneous scattering.

Summing over the number of detected photons leads to the total scattering probability $P_{\rm tot} = P_{\rm col} + P_{\rm sp}$ with the spontaneous scattering probability 
$P_{\rm sp} = \bar t_{\rm m} P_{\rm e} = \bar t_{\rm m} \bar I / (8 \bar\delta^2)$.  
We introduced dimensionless variables: time $\bar t_{\rm m} = \Gamma t_{\rm m}$ scaled by the natural linewidth $\Gamma$; detuning $\bar\delta = \Delta / \Gamma$ in units of $\Gamma$; and laser intensity $\bar I = I / I_{\rm sat}$ in units of the saturation intensity $I_{\rm sat}$. 
For our system $\Gamma/2\pi = 6.07\ {\rm MHz}$ and $I_{\rm{sat}} \approx 1.67~\rm{mW/cm^2}$.
It is convenient to parameterize this process in terms of a measurement strength $g=\sqrt{ \bar t_{\rm m} \bar I}/\bar \delta$ (giving $P_{\rm sp} = g^2/8$) quantifying the information extracted from the system by the hypothesized measurement process~\cite{Molmer1993, Altuntas2022_RIL}.

\section{Bolometry}\label{sec:Bolometry}

This section discusses how the energy deposited by the dispersive measurement of the atomic cloud is determined. 
First, we outline our bolometric experimental sequence for both a thermal gas and a BEC.
Then we present the details of our analysis procedure, which involves two primary steps. 
(1) We extract the temperature $T$ and the total number of atoms $N_{\rm t}$ from time-of-flight (TOF) resonant absorption images of the clouds, and (2) we then obtain the total energy $E_{\rm t}$ bolometrically using $T$ and $N_{\rm t}$.

\subsection{Experimental sequence}\label{sec:BolometrySequence}

Our protocol for bolometric measurements is as follows. 
The initial stage of each experimental sequence establishes a well-defined equilibrium state.
As shown in Fig.~\ref{Fig1:MeasurementModel}(b), we realize this by ramping down the trap depth [a crossed optical dipole trap (ODT) as will be detailed in Sec.~\ref{sec:ExpOverview}] and ensure the system is in thermal equilibrium characterized by a mean energy $E_0$ and with a corresponding temperature $T_0$.
The final value of the ODT power in this stage is varied to obtain a BEC or a thermal cloud, the two cases we studied. 
At this stage the equilibrium atomic system can be characterized by its entropy.

Next in the experimental sequence is a compression stage [Fig.~\ref{Fig1:MeasurementModel}(b)], where we increase the ODT power to a fixed value irrespective of $T_0$, i.e., the state of the atomic system.
This stage was designed to be adiabatic, i.e., isoentropic~\footnote{In practice we observe a reduction in the condensate fraction, which would not occur for a non-interacting Bose gas.}.
The compression procedure increases the trap depth, so that after adding energy evaporative processes do not reduce the temperature which would make bolometry ineffective. 

Following compression, the far-detuned probe is applied for a duration $t_m$ realizing a dispersive measurement characterized by a measurement strength $g$.
Immediately following the light-matter interaction, system is not in thermal equilibrium. 
For bolometry purposes, then we introduce a thermalization period during which the ODT power is kept constant to establish equilibration. 
Subsequently, the ODT is turned off, initiating time-of-flight. 
We detect the post-measurement atomic cloud using standard absorption imaging as illustrated in Fig.~\ref{Fig1:MeasurementModel}(b).
This completes our experimental sequence and in the following sections we review our analytical procedure for extracting the added-energy.

\subsection{Image analysis: number and temperature extraction}\label{sec:NT_analysis}

Standard absorption imaging begins with an image $I_A$ containing the shadow of the atomic ensemble in a large probe beam and a second image $I_P$ with the atoms absent.
The analysis commenced by computing the ratio of these two images $f = I_A / I_P$.
Both of these images contain diffraction fringes from dust and imperfections in the imaging system.
These imperfections can move on the wavelength scale in the time between the acquisition of the images, meaning that $f$ can contain spurious modulations from phase-shifted interference structures.
We use a principle component analysis (PCA) based technique to generate an ``optimal'' probe $I_{\rm PCA}$ for each $I_A$ to remove these artifacts.
In some data $f$ differs slightly from $1$ in regions where no atoms are present, giving an artificial background that we remove.
We then compute the $I_{\rm sat}$ corrected optical depth
\begin{align}
{\rm OD}^\prime = -{\rm ln}\left(\frac{I_A}{I_{\rm PCA}}\right) - \frac{I_A - I_{\rm PCA}}{I_{\rm sat}};
\end{align}
for further discussion see Ref.~\cite{Reinaudi2007}.

We extract temperature by excluding the central region (containing the Bose-condensed atoms) and performing a fit of the remainder to a 2D Gaussian model
\begin{align}
G(x,y) &= a_g \exp\left[-\frac{1}{2}\sum_{i=x,y}\left(\frac{x_i-b_i}{\sigma_i}\right)^2 \right],
\end{align}
where $a_g$ is the amplitude, $\sigma_{x,y}$ are the widths, and $b_{x,y}$ are center positions.
We implemented the exclusion by assigning extremely large uncertainties to data within the exclusion region.
The width of the excluded region along the $x$ and $y$ directions were set to be $20 \%$ and $15 \%$ larger than the largest observed Thomas-Fermi radius along that direction, respectively.
In our experiment, $\omega_y \approx 14.9\times  \omega_x$ so the excluded region was elliptical. 

In the fits, $\sigma_x$ and $\sigma_y$ vary independently, and as such we obtain two measures of temperature
\begin{align}
T_{x,y} = \frac{m}{\kb}\frac{\omega_{x,y}^2}{1+\omega_{x,y}^2 t^2}\sigma_{x,y}^2,
\end{align}
where $t$ is the TOF duration.
In our anisotropic trap (with $\omega_x \ll \omega_y, \omega_z$) the {\it in situ} extent of our clouds along $\ex$ was not small compared to the size in TOF.
As such $T_x$ has an $\approx 10\%$ correction compared to the long TOF limit whereas the correction for $T_y$ is negligible.
In addition, a quadruple magnetic field was present during TOF~\footnote{In a companion paper~\cite{Altuntas2022_RIL} this was utilized to separate the spin states via the Stern-Gerlach effect in other measurements using Ramsey interferometry to characterize back-action.}.
This quadruple field introduced curvature terms along $\ex$ reducing the width of the TOF distribution for our $\ket{F=2, m_F=2}$ ensembles.
In past experiments, this was observed to be an $\lesssim 5\ \%$ effect.
Since $T_y$ requires no correction factors, it would be favored as our primary measure of $T$.
However, we observed systematic shifts of $T_y$ (and not $T_x$) between different thermal gas measurement runs, for which $T_x$ and $T_y$ were not in agreement with each other, and as such we report $T = T_x$.

The number of atoms in thermal component is determined by integrating over the Gaussian profile, giving $N_{\rm nc} = 2 \pi \sigma_x \sigma_y a_g /\sigma_0$.
Here, $\sigma_0$ is the resonant scattering cross-section. 
To obtain the condensate population $N_{\rm c}$ we integrate the excluded region after subtracting the fitted thermal profile~\cite{Ketterle99, Szczepkowski2009}. 
Then the observed total number of atoms is $N_{\rm t} = N_{\rm c} + N_{\rm nc}$ and the condensate fraction is $R_{\rm c} = N_{\rm c} / N_{\rm t}$.

\subsection{Temperature-energy conversion}\label{sec:Temp_Energy}

We made complementary dispersive measurements in a dilute thermal gas and a BEC, and as such we evaluate temperature to energy conversion in both limits.
In the case of a thermal gas, energy is well approximated by the ideal gas result $E_{\rm t} = 3 N \kb T$ in a harmonic trap with Boltzmann constant $\kb$.

For a weakly interacting BEC the per-particle energy~\cite{Dalfovo1999} is
\begin{align}
\frac{E_{\rm t}}{N k_B T_{\rm c}^0} = \frac{3 \zeta(4)}{\zeta(3)}\bar T^4 +\frac{1}{7}\bar\mu(1-\bar T^3)^{2/5}(5+16\bar T^3),
\label{Eqn:TtoE}
\end{align}
in terms of $\bar T = T/T_{\rm c}^0$, $\bar\mu = \mu / (\kb T_{\rm c}^0)$, and the Riemann zeta function $\zeta(x)$.
Here, $T_{\rm c}^0$ is the 3D non-interacting BEC transition temperature.
In order to determine $T_{\rm c}^0$ we first extract the critical temperature $T_{\rm c}$ by fitting the observed condensate fraction $R_c$ to 
\begin{align}
R_c = \max\left[1 - \left(\frac{T}{T_{\rm c}}\right)^{a}, 0\right],
\label{Eqn:Tc}
\end{align}
where $T_{\rm c}$ and $a$ are fit parameters. 
Owing to the reduction of $T_{\rm c}$ with respect to the 3D non-interacting value, for the energy computation we use $T_{\rm{c}}^0$ acquired from our observed $T_{\rm c}$ following Eq.~(119) of Ref.~\cite{Dalfovo1999}: 
\begin{align}
\frac{\delta T_{\rm c}}{T_{\rm c}^0} = -1.3 \frac{a_{\rm{Rb}}}{a_{\rm{ho}}}N^{1/6},
\label{Eqn:Tc0}
\end{align}
where the shift in the critical temperature is $\delta T_{\rm c} = T_{\rm c} - T_{\rm{c}}^0$; $a_{\rm{Rb}}$ is the $^{87}$Rb scattering length; $a_{\rm{ho}} = (\hbar/ m \omega_{\rm{ho}})$ is the harmonic oscillator length with the geometric mean of trap frequencies $\omega_{\rm{ho}} = (\omega_x \omega_y \omega_z)^{1/3}$.
App.~\ref{app:T_toE} presents the measured critical temperature and the results for $T_{\rm c}^0$.  

\section{Experimental System}\label{sec:ExpOverview}

Our experiments started with ultracold $^{87}$Rb gases (both thermal and Bose-condensed) with about $1\times10^5$ atoms in the $\left|F = 1, m_{F} = 1\right\rangle$ electronic ground state in a crossed ODT. 
For the BEC case, ODT had trap frequencies $(\omega_x, \omega_y, \omega_z) = 2\pi \times \left[9.61(3), 113.9(3), 163.2(3) \right]\ {\rm Hz}$~\footnote{All uncertainties herein reflect the uncorrelated combination of single-sigma statistical and systematic uncertainties.}.
This trap configuration yielded condensates with condensate fraction $R_{\rm c} = 78(3)\%$, and chemical potential $\mu=h\times0.76(6)\ {\rm kHz}$.

As Sec.~\ref{sec:BolometrySequence} elaborated on, for the bolometric extraction of measurement-induced heating we then increased the trap depth yielding harmonic trap frequencies $(\omega_x, \omega_y, \omega_z) = 2\pi \times \left[22.6(3), 337(2), 265(2) \right]\ {\rm Hz}$. 
(We compressed by a reduced amount for data shown in Sec.~\ref{sec:Lattice}.)
Next, we applied a resonant microwave $\pi$ pulse to transfer the atoms into the $\ket{F=2,m_F=2}$ detection state.
We then implemented dispersive weak measurements by illuminating the BEC {\it in situ} with a far-detuned probe laser beam ($1/e^2$ radius $\approx 700~\mu\rm{m}$) on the $\ket{F=2, m_F=2}$ to $\ket{F^\prime=3,m_F^\prime=3}$ transition.
The measurement strength was adjusted by varying the probe laser detuning with $\bar\delta \in [-160 , 317]$, with intensity $\bar I$ up to 35, and the measurement time in the range $4~\mu\rm{s}< t_m < 20~\mu\rm{s}$.

The system was then allowed to thermalize for $400\ {\rm ms}$ [see Fig.~\ref{Fig1:MeasurementModel}(b)].
Subsequently, we extinguished the ODT and after a 20 ms TOF period detected the post-measurement density distribution using resonant absorption imaging ($20\ \mu{\rm s}$ pulse duration and intensity $I/I_{\rm{sat}}\approx 1$).
As previously noted, a Stern-Gerlach gradient was applied during TOF for consistency in experimental sequence with our companion paper~\cite{Altuntas2022_RIL}.
Using these TOF images we extracted the final temperature, and then the total energy $E_{\rm t}$ as discussed in Sec.~\ref{sec:Temp_Energy}.

Although, $\bar t_m$ and $\bar \delta$ are well determined by our experimental control sequence $\bar I$ is not.
We therefore imaged the dispersive measurement probe beam (with no atoms present) on a charge coupled device camera in order extract $\bar I$ (see Ref.~\cite{Altuntas2022_Isat}).
Together these allow us to compute $g$ with high accuracy.


\section{Light assisted collisions}\label{sec:Photoassiciation}

\begin{figure*}[tb!]
\begin{center}
\includegraphics{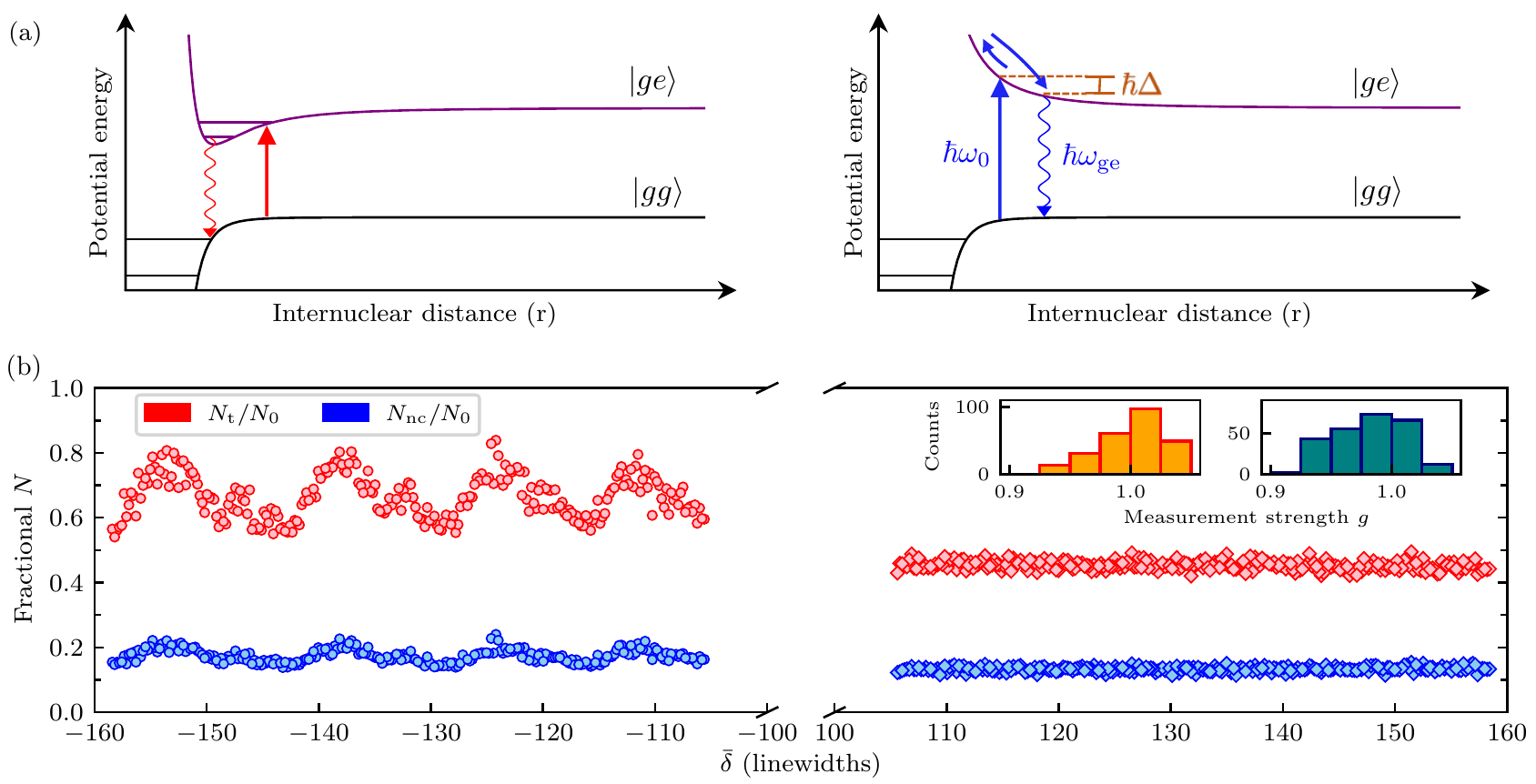}
\end{center}
\caption{
(a) Light assisted collisions for red (left) and blue (right) detuned light. 
Left: for red detuning atom pairs are excited to the attractive excited potential states followed by decay to ground state molecules. 
Right: for blue detuning atom pairs are excited to repulsive excited state and accelerate before decaying.
(b) Photoassociation losses following dispersive measurement.
Fraction of atoms remaining (red) and fraction remaining outside the BEC but within a 1 recoil momentum circle (blue).
The measurement consisted of two pulses each of duration $t_{\rm m}/2 = 8.2~\mu\rm{s}$ separated by a free-evolution time of $26.5~\mu\rm{s}$ following our pulse-evolve-pulse scheme described in Sec.~\ref{sec:Lattice}.
All data were taken with target measurement strength of $g = 1$ (attained by adjusting $\bar{I}$ between $15.5$ and $37$). 
The insets separately histogram the measurement strength sampled across red (left) and blue (right) detuned data.
}
\label{Fig5:PA_Losses}
\end{figure*}

Although the light matter interaction discussed in Sec.~\ref{sec:MeasurementModel} describes the behavior of atoms, it neglects the considerable impact of laser light on two-body molecular physics.
It is known from photoassociation experiments in cold atoms that near-resonant illumination can excite a zoo of molecular resonances~\cite{Cline1994}. 
In this section we present the results of such effects for dispersively measured BECs, and
App.~\ref{app:PCI_signal} details the PCI data resulting from these measurements.

Light assisted collisions describe enhanced 2-body collisions between atoms in the presence of a strong laser field~\cite{Jones_RevModP2006}.
Light assisted collisions manifest in two primary ways: losses from photoassociation (PA) and increased kinetic energy from light induced acceleration.
In our companion work~\cite{Altuntas2022_RIL}, we observed that probe light assisted collisions precipitate atom loss. 
This paper expands on the effect of such mechanisms in the context of measurement-induced heating.

The first mechanism, PA, is a 2-body loss process resulting from the formation of molecules.
Figure~\ref{Fig5:PA_Losses}(a) schematically illustrates Born-Oppenheimer scattering potentials for a pair of atoms both in the ground state (black) and with one excited atom (purple).
The potentials for red detuned light (left), include possible scattering resonances to electronically excited molecular states leading to PA. 
In a harmonic trap, PA preferentially removes atoms from regions of high density, located in the vicinity of the potential minima. 
As a consequence, this loss process increases the per-particle energy of the remnant atoms: anti-evaporation.

The second mechanism, light induced acceleration is illustrated by the potentials in Fig.~\ref{Fig5:PA_Losses}(a) for blue detuned light (right).
In this case, colliding atoms can acquire kinetic energy as they are promoted into the excited Born-Oppenheimer potential with a photon of energy $\hbar\omega_0$, but decay with a lower energy photon closer to $\hbar\omega_{\rm ge}$. 
This process directly adds kinetic energy, but conserves atom number.
In practice this process leads to loss by ejecting atoms from our comparatively shallow optical dipole trap.

Our {\it in situ} dispersive measurements were conducted  at high atomic densities of $\rho\approx1\times10^{14}\ {\rm cm}^{-3}$.
These experiments began with BECs with $N_0$ atoms in the $\ket{F=2,m_F=2}$ detection state; we then applied the far-detuned measurement light and immediately initiated TOF~\footnote{In this sequence, the compression stage was not employed and no Stern-Gerlach gradient was present during TOF.}. 
We determined the fractional change in total atom number $N_{\rm t}/N_0$ and in uncondensed number $N_{\rm nc}/N_0$. 
Here, $N_{\rm t}$ is computed by directly counting the atom number within a single photon recoil momentum circle centered on the BEC.
The fractional number $N_{\rm nc}/N_0$ has contributions from thermal atoms, those that have undergone large-angle light scattering, and atoms having undergone some light-assisted acceleration.

Figure~\ref{Fig5:PA_Losses}(b) plots these fractions as a function of $\bar\delta$ with constant $g=0.99(3)$, confirming the expected behavior for light assisted collisions.
The histograms present the distribution of measurement strengths for red and blue detuned data respectively, evidencing a nearly constant $g$ as $\bar \delta$ changes.
In the case of red detuning the fractional number is oscillatory, with minima marking the location of molecular resonances spaced by ``anti-resonances'' with reduced loss.
By contrast, the blue detuned data is completely featureless.

The observed peak atom number at the anti-resonances is nearly $2\times$ larger than that is for blue detuning. 
However, the uncondensed fraction is only slightly increased.
This likely results from light-assisted collisions occurring predominantly in the high density BEC, rather than the surrounding lower density thermal cloud.

A second important feature of Fig.~\ref{Fig5:PA_Losses}(b) is that for both red and blue detuning the nominal loss rate at fixed $g$ has no overall dependence on $\bar \delta$. 
Instead $\bar \delta$ serves only to control the detuning from molecular resonances.
This results from the fact that the overall rate of light assisted collisions is proportional to the excited state probability $P_{\rm e}$, yielding a collision number in the time interval $t_{\rm m}$ proportional to $g^2$.
Importantly this is the same scaling behavior as for light scattering.

\section{Heating Measurements} \label{sec:MI_heat_results}

The back-action resulting from the environment detecting each scattered photon invariably adds energy to the system as $\ket{\Psi_{\rm S} } \rightarrow \ket{\Psi_{\rm S}^\prime}$.
The added energy is an extensive quantity resulting from the change in $\ket{\Psi_{\rm S}^\prime}$ with contributions both from kinetic and interaction energies.
In BECs, the added interaction energy derives from the change in density as outgoing scattered atoms interfere with the BEC mode.
We obtain the total energy $E_{\rm t}$ bolometrically by measuring the temperature $T$ following thermalization, and contrast the cases of a BEC and a thermal gas.

\subsection{Compressed BEC heating}\label{sec:Comopressed_BEC_heating}

\begin{figure}[htb!]
\includegraphics{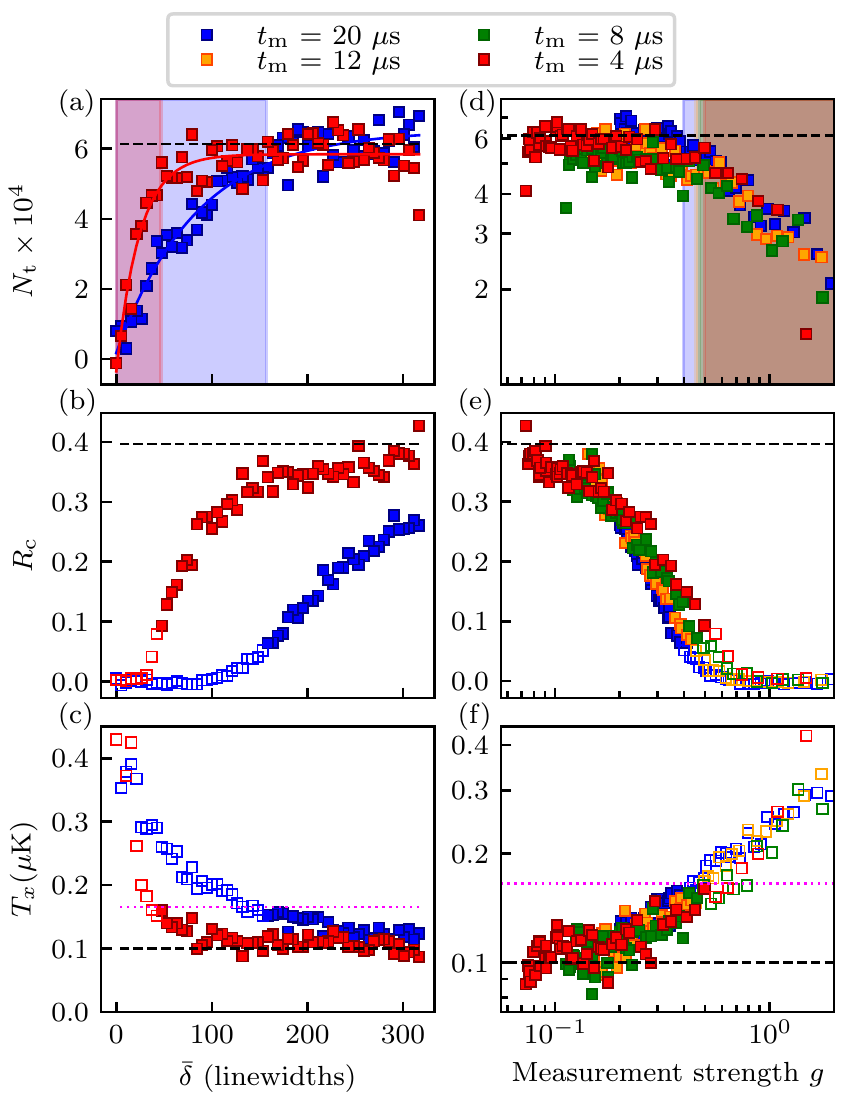}
\caption{
Heating of a BEC dispersively-measured with a single probe pulse.
Left: Results for measurement times $t_{\rm {m}} = 4~\mu\rm{s}$ and $20~\mu\rm{s}$ plotted as a function of $\bar\delta$.
Right: Data plotted as a function of $g$ on logarithmic scale, with data excluded as described in the text plotted with hollow symbols.
(a), (d) Total number $N_{\rm {t}}$. 
The solid curves in (a) are fits to Eq.~\eqref{eq:num_fit} and the shaded regions are where $N_{\rm {t}}$ has fallen below $85\ \%$ of its asymptotic value.
(b), (e) Condensate fraction $R_{\rm c}$.
(c), (f) Temperature $T$.  
The magenta dashed line marks the condensation temperature $T_c$.
In each panel, the dashed black lines indicate the results when no-dispersive measurement took place.
Each data point is the average of 5 iterations of the experiment. 
}
\label{Fig5:BEC_NRT}
\end{figure}

Our data generally consisted of bimodal density distributions with contributions from condensed and uncondensed atoms. 
We obtain $T$ as well as $N_{\rm c}$ and $N_{\rm nc}$ as described in Sec.~\ref{sec:NT_analysis}.
Figure~\ref{Fig5:BEC_NRT}(a) shows the total number $N_{\rm t}$ obtained via this procedure as a function of $\bar\delta$ for two measurement times $t_{\rm m}$.
The variation in number for different pulse times at large $\bar\delta$ derives from long-term number drift in our apparatus.
These data show that for sufficiently small detuning of the probe beam $N_{\rm t}$ begins to decrease potentially due to evaporation and  photoassociative losses (discussed in Sec.~\ref{sec:Photoassiciation}).
We therefore treat $N_{\rm t}$ as a gate marking data for which bolometry is valid; the curves depict fits to 
\begin{align}
f = N_0-A \exp[-B /  |g(t_m, \bar\delta, \bar I)|],
\label{eq:num_fit}
\end{align}
(a falling exponential function of $\bar\delta$) and we accepted data when the curve exceeds $85\ \%$ of its maximum value (unshaded regions).
In the following figures data rejected by this threshold are plotted as hollow symbols.

The condensate fraction $R_{\rm c}$ and temperature $T$ shown in Fig.~\ref{Fig5:BEC_NRT}(b) and (c) consistently indicate that increasing the measurement strength parameter $g$---either from reducing $\bar\delta$ or increasing $t_{\rm m}$---are increasingly destructive, increasing $T$ and reducing $R_{\rm c}$ in tandem. 
Using the procedure described in App.~\ref{app:T_toE}, the condensate fraction in (b), in conjunction with the temperature in (c), allow us to extract the BEC transition temperature $T_{\rm c} = 165\ {\rm nK}$ for our $N\approx10^5$ atom system [horizontal magenta dotted lines in (c) and (f)], which is reduced with respect to the 3D non-interacting value of $T_{\rm c}^0 = 225\ {\rm nK}$ for a harmonically trapped gas.
This suppression primarily results from our system being transitionary from 3D to 1D with $\mu$ only about $3$ times larger than the transverse trap frequencies. 
Panels (d)-(f) expand on this observation by plotting such data taken for four values of $t_{\rm m}$ as a function of $g$ and show all of our observations nominally collapse onto the same curve.

\subsection{BEC heating with stray lattice mitigated}\label{sec:Lattice}

As we shall see in Sec.~\ref{sec:AddedEnergy}, the energy added by these dispersive measurements is greatly in excess of that predicted by our scattering model.
We found that in our experiment, the probe beam creates a stray optical lattice by interfering with its retro-reflections off the optical elements in the high-resolution {\it in situ} imaging system.
In optical setups, it is a standard practice to introduce slight tilts in optical elements to prevent back-reflections. 
On the other hand, in a high-resolution imaging setup the probe beam is centered on the optical axis and optimized to intersect each element at normal incidence in order to minimize optical aberrations ~\cite{Altuntas2021}.
Consequently, in our experiment a weak optical lattice is generated as a systematic byproduct during each dispersive-measurement probe pulse. 
Matterwave diffraction of the BEC off of a weak optical lattice coherently creates population in diffraction orders with momentum $\pm2\hbar \kr$.

The phase imprinted by the stray lattice can be unwound by splitting the probe pulse into two pulses of shorter duration $t_{\rm p}$ separated by a delay time of $t_{\rm d}$ (see in our recent work in Ref.~\cite{Altuntas2022_RIL} for a more detailed discussion).
Our technique for mitigating the optical lattice can be intuitively understood in terms of a three-state truncation~\cite{Wu2005,Trey_2012PRL} of the full lattice Hamiltonian
\begin{align}
\frac{\hat H(k)}{\Er} &= 
\left(\begin{array}{ccc}
(k+2\kr)^2 & s/4 & 0 \\
s/4 & k^2 & s/4 \\
0 & s/4 & (k-2\kr)^2
\end{array}\right),
\end{align}
describing a lattice of depth $s\Er$, with single photon recoil momentum $\hbar\kr = 2\pi\hbar/\lambda$, energy $\Er=\hbar^2 \kr^2 / (2 m)$, and time $T_0 = 2\pi\hbar/\Er\approx 265\ \mu{\rm s}$.
For atoms initially at rest, i.e. $k=0$, this is a resonant lambda coupling scheme with bright state subspace spanned by $\ket{b_0}=\ket{k=0}$ and $\ket{b_1}=(\ket{k=-2\kr}+\ket{k=-2\kr}) / \sqrt{2}$ and an uncoupled dark state $\ket{d}=(\ket{k=-2\kr}-\ket{k=-2\kr}) / \sqrt{2}$.

Our initial state $\ket{k=0}$ is in the bright state manifold, so we focus on the bright state Hamiltonian
\begin{align*}
\frac{\hat H_{\rm b}(0)}{\Er} &= 2 \hat I + \frac{1}{2}\left[4\hat \sigma_z + \frac{s}{\sqrt{2}} \hat \sigma_x\right].
\end{align*}
When the lattice is off, this describes Larmour procession around $\ez$ with frequency $4\Er/\hbar$ and when the lattice is on the procession axes changes to $4 \ez + [s / \sqrt{2}]\ex$ with frequency $\sqrt{16 + s^2/2} \Er/\hbar$.
In the limit $s\ll4\sqrt{2}$, the axis of rotation is tipped by $\theta = 4s/\sqrt{2}$, the Rabi frequency is nearly unchanged from $4\Er/\hbar$, and the condition to return to the initial state is $t_{\rm d}/T_0 = 1/8 - t_{\rm p}/T_0$.
In practice we selected $t_{\rm d} = T_0/10 = 26.5\ \mu{\rm s}$ and $t_{\rm p} = T_0/32 = 8.2\ \mu{\rm s}$.

This pulse-evolve-pulse scheme is only effective for momenta near zero, thereby rendering it ineffective for measurements at higher temperature.
These include data from the thermal cloud as well as the BEC in the deep trap, for which $R_{\rm c}\approx 0.4$ [see Fig.~\ref{Fig5:BEC_NRT}(b) and (e)].
For this reason, we altered our experimental sequence to make measurements with the pulse-evolve-pulse sequence at reduced temperature (with $T = 41\ {\rm nK}$ and $R_{\rm c}\approx 0.77$).
We reduced the temperature by first starting with a colder BEC [$R_{\rm c}\approx 0.97(3)$] and then increasing the trap depth only to $\approx 3\times \Er$ (a factor of $\approx 3.3$ shallower than the data discussed above) yielding final trap frequencies of $(\omega_x, \omega_y, \omega_z) = 2\pi \times \left[13.1(1), 206.7(8), 214.3(5) \right]\ {\rm Hz}$. 
This gives reduced heating due to compression, and is still sufficient to trap scattered atoms and avoid evaporation.


\subsection{BEC heating at PA anti-resonances}\label{sec:BEC_Red}

Having eliminated excess heating from lattice effects, we now turn to light assisted collisions.
The heating data presented to this point was from dispersive measurements with the probe light blue detuned from resonance.
As we discussed in Sec.~\ref{sec:Photoassiciation} light assisted collisions in this regime depend only on excited state probability, and otherwise are independent of $\bar\delta$.
By contrast PA resonances and anti-resonances are present for red detuned probe light.

\begin{figure}[hbt!]
\includegraphics{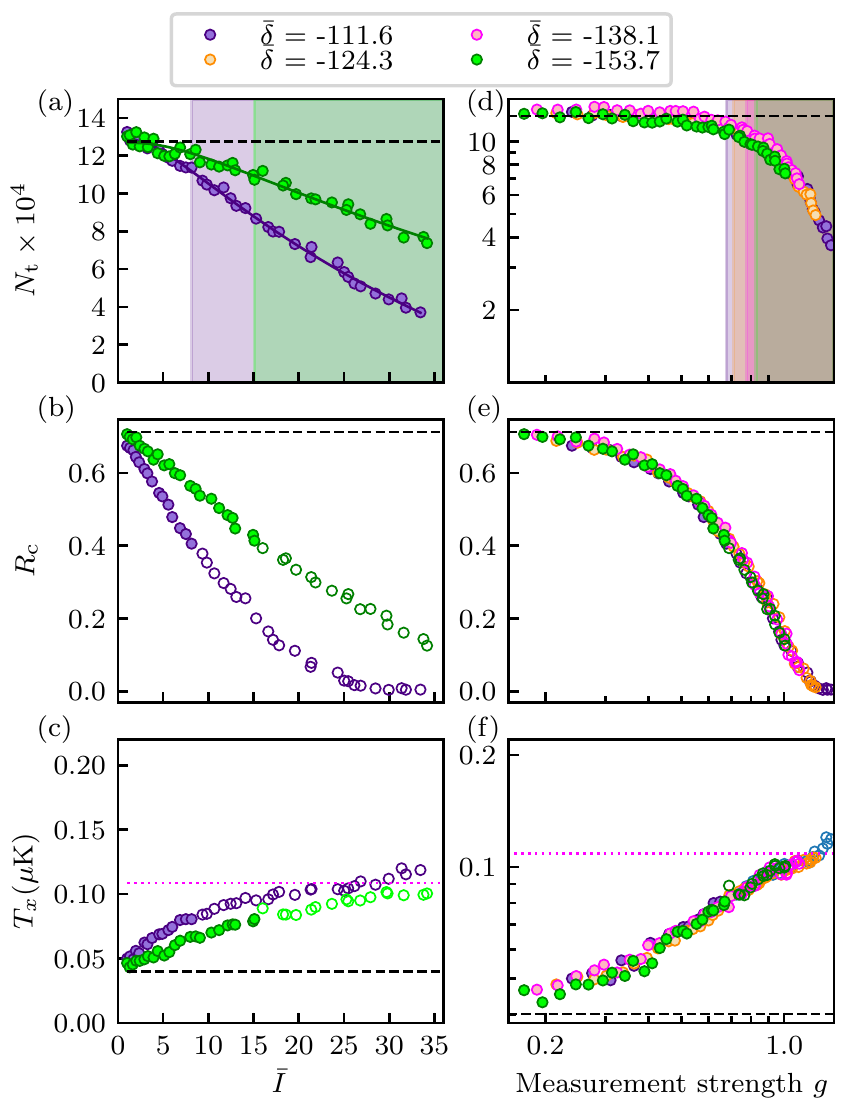}
\caption{
Heating in a BEC dispersively-measured at PA anti-resonances and with stray lattice mitigated using the pulse-evolve-pulse scheme. 
Left: Results for detunings $\bar\delta = -111.6$ and $\bar\delta= -153.7$ plotted as a function of $\bar I$.
Right: Data for all four $\bar\delta$ values plotted as a function of $g$ on logarithmic scale.
(a), (d) Total number $N_{\rm {t}}$. 
The solid curves in (a) are fits Eq.~\eqref{eq:num_fit} and the shaded regions indicate where $N_{\rm {t}}$ falls below $85\ \%$ of its asymptotic value. 
(b), (e) Condensate fraction $R_{\rm c}$.
(c), (f) Temperature $T$.  
Hollow symbols indicate excluded data in added-energy computation as in Fig.~\ref{Fig5:BEC_NRT}.
In each panel, the dashed black lines indicate the results when no-dispersive measurement took place.
Each data point is the average of 5 iterations of the experiment.
} 
\label{Fig:RedDetuned_N_R_T}
\end{figure}

We concentrate our measurements on the anti-resonant features observed at $\bar\delta\in\{-111.6,$ $-124.3,$ $-138.1,$ $-153.7\}$ and to further mitigate excess heating continue to use the pulse-evolve-pulse sequence.
As such, the intensity $\bar I$ is the only remaining parameter by which the measurement strength $g$ can be tuned.
Figure~\ref{Fig:RedDetuned_N_R_T} shows the results of these measurements with panels (a)-(c) containing curves taken at two anti-resonances.
As with previous analysis, we retain data where the number has dropped by less than 15\ \% for obtaining the deposited energy.
Fig.~\ref{Fig:RedDetuned_N_R_T}(d)-(f) demonstrate that the same data plotted as a function of $g$ nearly perfectly collapse.

In Secs.~\ref{sec:AddedEnergy} and \ref{sec:discussion} we discuss and contrast the data sets presented thus far.


\subsection{Thermal gas heating}\label{sec:ThermalGas}

As a reference case we also measured the change in temperature of a dilute thermal gas at $T/T_{\rm c} \approx 2.5$.
To facilitate comparison with BEC measurements, we used parameters common with the BEC data: blue detuned thermal data was taken following the procedure in Sec.~\ref{sec:Comopressed_BEC_heating}, while red-detuned thermal data followed Sec.~\ref{sec:BEC_Red}.
These thermal data used a single probe pulse of duration $t_m = 20~\mu\rm{s}$ (we did not apply the pulse-evolve-pulse sequence to these higher temperature data as it is ineffective in this regime).
Appendix~\ref{app:ThermalGas} presents the extracted $N_t$ and $T$ at red detuned probe light measurements.
Similar to the BEC measurements reported in Figures~\ref{Fig5:BEC_NRT} and \ref{Fig:RedDetuned_N_R_T}, these data demonstrate that $N_t$ and $T$ collapse when scaled to $g$.
In the next section, we compare the added energy in these thermal gas measurements with the ones in BECs.

\subsection{Added energy}\label{sec:AddedEnergy}

\begin{figure}[t]
\includegraphics{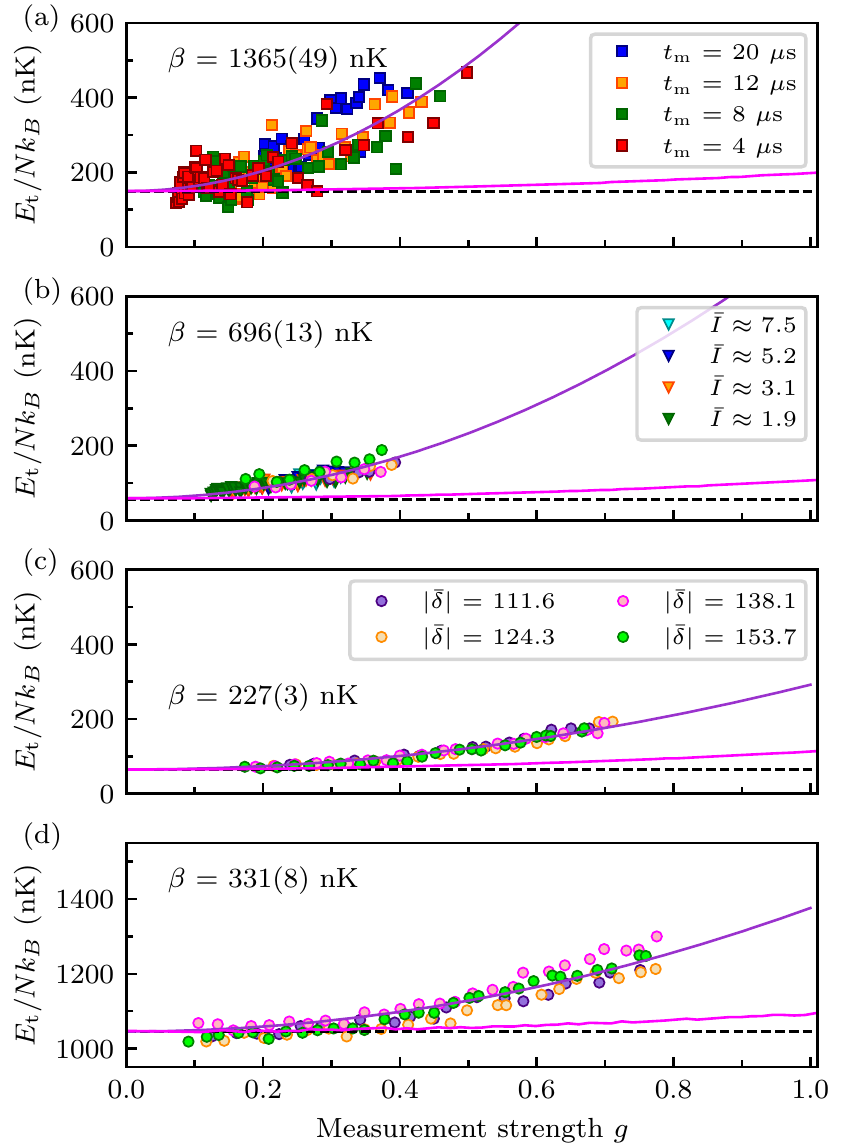}
\caption{
Per-atom added energy.
(a) BEC single probe pulse measurements at blue detuning.
(b) BEC pulse-evolve-pulse scheme measurements at blue detuning.
(c) BEC pulse-evolve-pulse scheme measurements at PA anti-resonances.
(d) Thermal gas single probe pulse measurements at PA anti-resonances.
Squares indicate data for which $\bar\delta$ was scanned at four values of $t_{\rm m}$, and with $\bar I$ held constant.
Triangles mark data for which $\bar \delta$ was scanned at four values of $\bar I$, at constant $t_{\rm m}$.
Circles indicate data for which $\bar I$ was scanned at four values of $\bar\delta$, at constant $t_{\rm m}$.
The squares in (a) and triangles in (b) are jointly colored according to $\sqrt{\bar t_{\rm m} \bar I}$, the numerator of the measurement strength expression.
The circular symbols in (b), (c) and (d) use the common legend in (c) identifying different values of $|\bar\delta|$.
The magenta curves plot the predicted added kinetic energy obtained from Monte Carlo simulations.
The purple curves are fits to $E = E_0 + \beta g^2$ and the black dashed lines depict the initial energy.
}
\label{Fig4:Energy}
\end{figure}

Having measured the temperature both of condensed and thermal systems, we now consider the per-atom energy~\footnote{We calculate the per-atom added energy instead of the total energy to eliminate the correlated uncertainty present in total energy (an extensive quantity) resulting from the uncertainty in $N_{\rm t}$.} using the conversions described in Sec.~\ref{sec:Temp_Energy}.

Using the data presented in Sec.~\ref{sec:Comopressed_BEC_heating}, Fig.~\ref{Fig4:Energy}(a) shows $E_{\rm t}$ for a weakly interacting BEC (markers) along with a fit to the expected functional form $\delta E = E_0 + \beta g^2$. 
The magenta curve plots the energy computed from a stochastic classical scattering model (see App.~\ref{app:ScatModel}), which includes only large-angle scattering, i.e., spontaneous emission~\footnote{
This prediction neglects the increased interaction energy, which even for nearly pure condensates contributes a per-atom energy of $\approx\mu$ per scattering event, increasing $\beta$ by a negligible $\approx 4.5\ {\rm nK}$. 
}; this gives $\beta = 54.9(3)\ {\rm nK}$ as compared to the fit value $\beta = 1365(49)\ {\rm nK}$. 

Figure~\ref{Fig4:Energy}(b) incorporates the pulse-evolve-pulse measurement protocol in which we varied $\bar\delta$ at four different probe intensities $I$ keeping $t_{\rm {m}}$ constant [the marker colors in (b) are selected to make the product $\bar t_{\rm m} \bar I$ consistent with the colors in (a).]
We find that the added energy $E_{\rm t}$ is decreased by half compared to the single pulse measurements, giving $\beta=696(13)\ {\rm nK}$.
While this is a marked improvement it is still more than ten times in excess of the simple spontaneous emission prediction.

Lastly, Fig.~\ref{Fig4:Energy}(c) retains the pulse-evolve-pulse protocol and operates at the PA anti-resonances.
As discussed in Sec.~\ref{sec:BEC_Red}, for these measurements we held $t_{\rm {m}}$ and $\bar\delta$ constant and scanned $\bar I$.
In these data the heating rate is reduced by a factor of about three giving $\beta=227(4)\ {\rm nK}$.
This represents a $6\times$ reduction in heating as compared to the blue detuned data in (a) without lattice compensation, however, it is still about $4.1\times$ in excess of our expectations.

By comparison, Fig.~\ref{Fig4:Energy}(d) presents thermal gas data (at red-detuned anti-resonances) for which we obtain a fit coefficient $\beta = 331(8)\ {\rm nK}$ and a similar analysis for blue detuning yields $\beta = 401(8)\ {\rm nK}$.
In contrast with BEC measurements, compensating for light assisted collisions yielded a modest 20\% improvement.
This is expected owing to the greatly reduced density of the thermal gas.

\subsection{Discussion}\label{sec:discussion}

In all cases the added energy is far in excess of what is expected from light scattering alone.
The results presented in Fig.~\ref{Fig4:Energy} confirm that both the stray lattice and light assisted collisions contribute.

The energy added by the stray optical lattice, a function of its depth $V_{\rm latt}\propto \bar I / \delta$ as well as the time $\bar t_m$, has no particular relation with the scattering probability $g^2/8$.
As a result lattice induced heating does not yield data collapse when scaled to $g$ (see App.~\ref{app:ScatModel} for numerical examples).
The single-pulse BEC measurements data shown in Fig.~\ref{Fig4:Energy}(a) collapses poorly; in conjunction with the energy reduction and improved collapse of the pulse-evolve-pulse data presented in Fig.~\ref{Fig4:Energy}(b) and (c), this is fully consistent with the stray lattice as a significant contributor of BEC heating.
On the other hand, data from the single-pulse measurements in thermal gas in Fig.~\ref{Fig4:Energy}(d) show reasonable collapse, with the imperfect collapse being consistent with the simulations in App.~\ref{app:ScatModel} (for which the peak lattice depth was about $5\Er$).

In comparison, the light assisted collision rate is proportional to the excited state occupation probability, so the short-time rate of such collisions is $\propto g^2$.
Therefore they contribute heating with the same overall scaling as spontaneous scattering.
As such, the observed progression from Fig.~\ref{Fig4:Energy}(a) to (c) is consistent with heating from a combination of photon scattering and light assisted collision processes.

In addition Fig~\ref{Fig4:Energy}(c) and (d), show that the added energy for the BEC is smaller than that of the thermal gas.
This in line with our expectations because lattice compensation is ineffective for the broad momentum distribution of a thermal gas.
In the case of the fully compensated BEC, it is unclear whether the remaining excess energy results from imperfect cancellation of the lattice, or other effects.
Although expected to be a minor effect, the probe beam inhomogeneities, which were experimentally characterized for our specific setup in Ref.~\cite{Altuntas2022_Isat} will impart some energy.

\section{Conclusion and outlook}
\label{sec:Conc}

In this paper, we characterized heating of dispersively measured ultracold atoms and identified systematic effects that dominate the heating with respect to the quantum back-action signal. 
Nevertheless, straightforward applications of dispersive imaging techniques may well allow repeated monitoring of the same quantum system.
In this case, excess heating places additional limitations on the lifetime of continuously monitored BECs, and further constrains on potential applications.
In a complementary measurement with the same experimental setup we found that the reduction in contrast of a Ramsey interferometer is back-action limited~\cite{Altuntas2022_RIL}.
This apparent contradiction indicates that not all degrees of freedom are equally impacted by these systematic effects.

The systematic effects---a stray optical lattice and light-assisted collisions---accentuate critical requirements for future research with back-action limited measurements.
A feasible experimental modification to minimize the systematic contribution of stray optical lattices would use a probe beam slightly tilted with respect to the existing probe beam.
As noted in Sec.~\ref{sec:Lattice}, high-resolution cold-atom imaging systems generally use the probe beam to define the optical axis, allowing the optical elements to be placed on-center and at normal incidence with respect to the optical axis.
This makes a well-aligned probe beam indispensable for imaging system alignment.
However, this reference probe beam need not be used for actual imaging: a second, slightly tilted, probe could greatly mitigate back reflections, at the price of potentially increased aberrations.
Thus the original perfectly-centered probe beam functions to define the optical axis, while the second tilted probe beam performs the dispersive measurements and governs the quantum back-action.
The performance of our pulse-evolve-pulse scheme improves for weaker lattices, so modestly reducing back reflections can yield disproportionate benefits.

Secondly, light-assisted collisions can be suppressed in BECs either by controlling the atom density, or by further management of molecular resonances~\cite{Urvoy2019}.
For example homogeneous confining potentials, i.e. ``box traps'', reduce $\overline{\rho^2}$ the average value of density squared, and therefore decrease the rate of two-body effects such as light assisted collisions.
In addition the spacing between molecular resonances increases with detuning~\cite{Cline1994}, allowing for more robust anti-resonances at large detuning.

There are multiple measurement techniques for quantum gases based on the dispersive light-matter interaction~\cite{Andrews1996, Ketterle99, Anderson2001, Gajdacz2013} that in principle can give back-action limited measurement outcomes.
In particular, we use digitally enhanced phase-contrast imaging~\cite{Altuntas2021}, an optical homodyne detection technique that accounts for imaging system imperfections.
For example, App.~\ref{app:PCI_signal} presents PCI data associated with the light assisted collisions measurements in Fig.~\ref{Fig5:PA_Losses}(b).
The second-order light-matter interaction in Eq.~\eqref{Eqn:acStark_Hamiltonian} is $\propto 1/\delta$ and purely dispersive in the classical limits:  when $\hat E\rightarrow\langle \hat E\rangle$ the atoms experience only an AC Stark shift, and when $\hat n_{\rm g}\rightarrow\langle \hat n_{\rm g}\rangle$ the light experiences only a phase shift.
Despite this,  Eq.~\eqref{Eqn:acStark_Hamiltonian} accounts for both spontaneous and stimulated emission with their $\propto 1/\delta^2$ and $\propto 1/\delta$ scaling respectively.
This gives both the imaginary (dissipative) and real (dispersive) parts of the atomic susceptibility. 
As such this theoretical approach can be applied even quite close to resonance, until these scalings break down.

The $4\pi$ steradian measurement model outlined in this paper is powerful and as demonstrated in our companion paper Ref.~\cite{Altuntas2022_RIL} makes reliable predictions.
While a convenient theoretical abstraction, this model does not derive from a practical experimental measurement geometry.
By contrast measuring the spatially resolved optical phase shift via PCI is a well-established dispersive measurement method.
However, in this case the conditional change in the system wave function given an observed real space measurement is not described by the intuitive picture of atoms recoiling from scattered photons.
Formally, this measurement is associated with a completely different set of Kraus operators, corresponding to a different (and physically motivated) unraveling of the master equation.
A natural extension of this work will account for these differences and explore the experimental applications. 

\begin{acknowledgments}
This work was partially supported by the National Institute of Standards and Technology, the National Science Foundation through the Quantum Leap Challenge Institute for Robust Quantum Simulation (grant OMA-2120757), and by the Air Force Office of Scientific Research's Multidisciplinary University Research Initiative ``RAPSYDY in Q'' (grant FA9550-22-1-0339).
\end{acknowledgments}

\appendix

\section{BEC Thermodynamics}\label{app:T_toE}

\begin{figure}[tb!]
\includegraphics{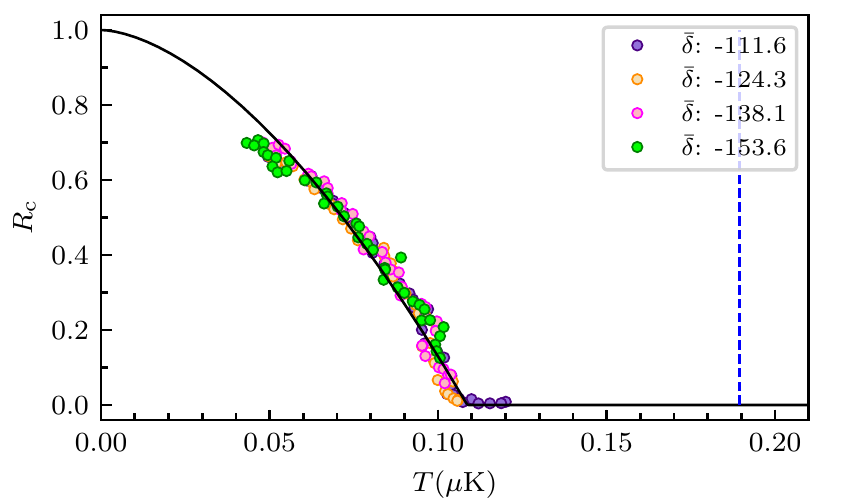}
\caption{Determination of critical temperature. 
The vertical blue dashed line marks the predicted $T_{\rm c}^0$ for non-interacting bosons computed from for our trap frequencies in the compressed trap and total atom number $N_{\rm{nm}}$. 
}
\label{FigSM:BEC_Tc_extract}
\end{figure}

The per-particle energy for a weakly interacting BEC given in Eq.~\eqref{Eqn:TtoE} requires the knowledge of two critical experimental parameters: the 3D non-interacting BEC transition temperature $T_{\rm c}^0$ and the chemical potential $\mu$. 
This section provides the analysis details for obtaining each parameter. 

We determine $T_{\rm c}^0$ using the measured critical temperature $T_{\rm c}$ as detailed in Sec.~\ref{sec:Temp_Energy}.
Figure~\ref{FigSM:BEC_Tc_extract} presents $R_c$ as a function of $T$ using the BEC heating at anti-resonances data presented in Sec.~\ref{sec:BEC_Red}.
Fitting to Eq.~\eqref{Eqn:Tc} gives best fit results $T_{\rm c} = 109(1)\ {\rm nK}$ and $a = 1.67(1)$. 
For the no-weak-measurement BEC results (dashed lines in Fig.~\ref{Fig5:BEC_NRT}) we determine the number of atoms to be $N_{\rm{nm}} = 1.28(1)\times10^5$. 
For our compressed ODT configuration, Eq.~\eqref{Eqn:Tc0} yields $T_{\rm c}^0 = 189\ {\rm nK}$. 

According to the Thomas-Fermi approximation, the chemical potential is related to the condensed atom number by 
\begin{align}
\mu = \frac{\omega_{\rm{ho}}}{2} \left[\frac{15 N a_{\rm{Rb}}}{a_{\rm{ho}}}\right]^{2/5}.
\label{Eqn:mu_BEC}
\end{align}
We derive the chemical potential $\mu$ using $N_{\rm{nm}}$ value with the measured compressed trap frequencies.

\section{Phase-Contrast Imaging} \label{app:PCI_signal}

\begin{figure*}[htb!]
\begin{center}
\includegraphics{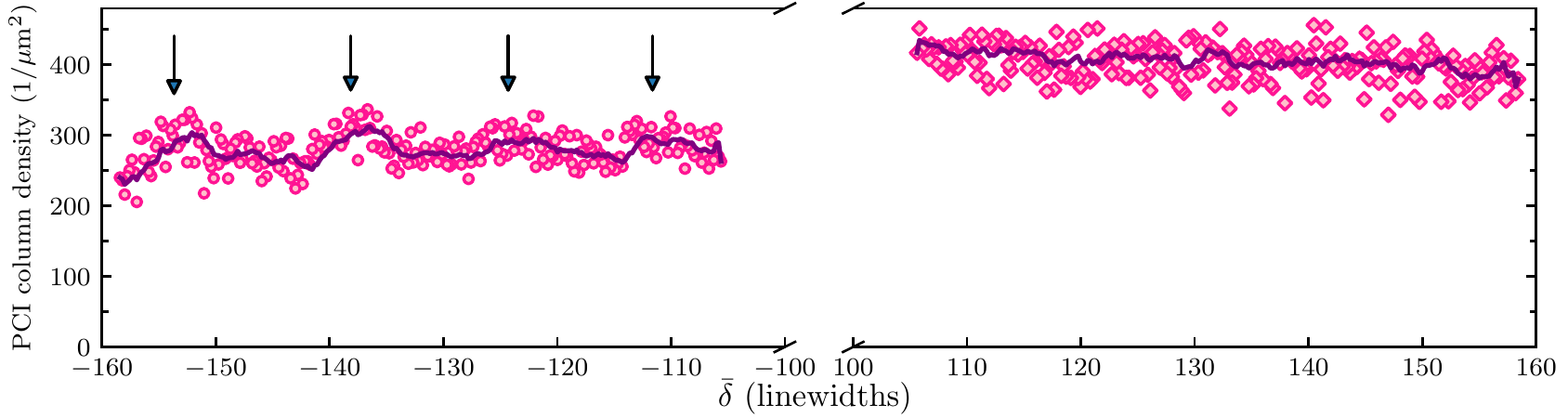}
\end{center}
\caption{
Detected peak atomic column density computed from PCI data for red (left) and blue (right) detuning. 
These data result from the exact same dispersive measurements that lead to the photoassociation and light assisted acceleration shown in Fig.~\ref{Fig5:PA_Losses}.
The purple curves plot a 10 point wide running average as a guide to the eye.
The vertical arrows mark the PA anti-resonances observed in Fig.~\ref{Fig5:PA_Losses}(b).
All data were taken with target measurement strength of $g = 1$ (attained by adjusting $\bar{I}$ between $15.5$ and $37$). 
}
\label{FigApp:PCI_signal}
\end{figure*}

Although the measurement strength $g$ was introduced in the context of a $4\pi$ scattering measurement model, it quantifies the strength of any measurement based on the dispersive light-matter interaction.
For example, the far-detuned probe laser described here actually implements PCI in our laboratory, and the PCI signal to noise ratio is proportional to $g$.
Although it is not the central focus of the present work, we did acquire a phase-contrast image each time we illuminated the BEC with the probe laser; as such the probe laser truly implemented dispersive measurements.
Details regarding our PCI setup can be found in Refs.~\onlinecite{Altuntas2021} and \onlinecite{Altuntas2022_Isat}.

Figure~\ref{FigApp:PCI_signal} plots the {\it in situ} peak density determined from PCI (see Ref.~\onlinecite{Altuntas2021} for details) for each TOF data point presented in Fig.~\ref{Fig5:PA_Losses}.
In both panels the solid curves plot the data averaged over 10 points and the arrows mark the location of the PA anti-resonances observed in Fig.~\ref{Fig5:PA_Losses}.

The PCI and TOF data are different in three qualitative ways:
(1) the signal to noise ratio of the PCI data is reduced;
(2) the contrast of the red detuned PA features is reduced; and (3) the PCI signal is larger for the blue detuned data, while the reverse is the case after TOF.
Firstly, the PCI signal to noise ratio is reduced by about $4\times$ compared to that in TOF simply because PCI is a weak measurement process.

Second, the red detuned PA features are reduced in amplitude because the PCI signal is proportional to the atom number averaged over the measurement pulse's duration, not the number following the pulse as for TOF data.
When the loss is small, this argument predicts a $50\ \%$ reduction in contrast.
An additional contributor is remnant absorption at large detunings; this impacts the PCI signal in a way that is anti-symmetric in detuning.
For a phase dot with $-\pi/2$ phase shift as in our case, absorption increases the PCI signal for blue detuning and reduces it for red detuning.
This introduces an additional $\approx 15\ \%$ fractional difference between the red and blue detuned data.

The third observation results from the fact that the light induced acceleration process does not change the atom number, and during our brief measurement pulses the accelerated atoms do not have sufficient time to leave the BEC. 
By contrast the red detuned PA process simply removes atoms.
This observation also indicates that individual PCI measurements give a misleading impression regarding the importance of light assisted collisions. 

Considering only the direct effect of the atoms on the
far-detuned probe light, without quantifying the resulting heating of the post-measurement cloud, cloaks systematic effects of light-assisted collisions. 
Indeed, our heating measurements (see Fig.~\ref{Fig4:Energy}) show that the blue-detuned process alone adds about 8 times more energy than can be attributed to the light matter interaction alone.
Each of these observations gives different and important facts informing the design of experiments focused on measurement back-action.

\section{Thermal Gas Measurements} \label{app:ThermalGas}

\begin{figure}[tb!]
\includegraphics{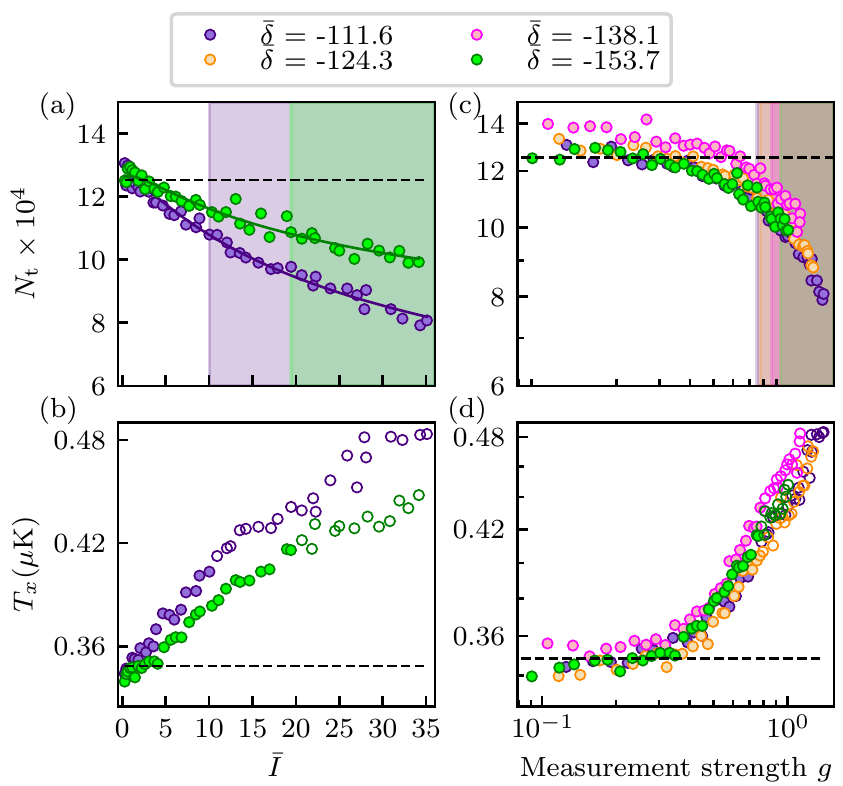}
\caption{
Heating of a thermal gas dispersively measured at PA anti-resonances.
Left: Results for probe detunings $\bar\delta = -111.6$ and $\bar\delta= -153.7$ plotted as a function of $\bar I$.
Right: Data for all four $\bar\delta$ values plotted as a function of $g$ on logarithmic scale.
(a), (c) Total number $N_{\rm {t}}$. 
The solid curves in (a) are fits Eq.~\eqref{eq:num_fit} and the shaded regions show where $N_{\rm {t}}$ falls below $85\ \%$ of its asymptotic value. 
(b), (d) Temperature $T$.   
All measurements were of a single probe pulse of duration $t_{\rm {m}} = 20~\mu\rm{s}$.
Hollow symbols mark excluded data in added-energy computation described in the main text.
In each panel, the dashed black lines indicate the results when no-dispersive measurement took place.
Each data point is the average of 5 iterations of the experiment.
}
\label{FigSM:ThermalGas_N_T}
\end{figure}

For completeness, Fig.~\ref{FigSM:ThermalGas_N_T} displays the data acquired from dispersive measurements in a dilute thermal gas.
These results were then used to arrive at the data presented in Fig.~\ref{Fig4:Energy}(d) for post-measurement energy increase in a thermal gas.

\section{Heating from anisotropic scattering} \label{app:ScatModel}

\begin{figure}[tb!]
\includegraphics{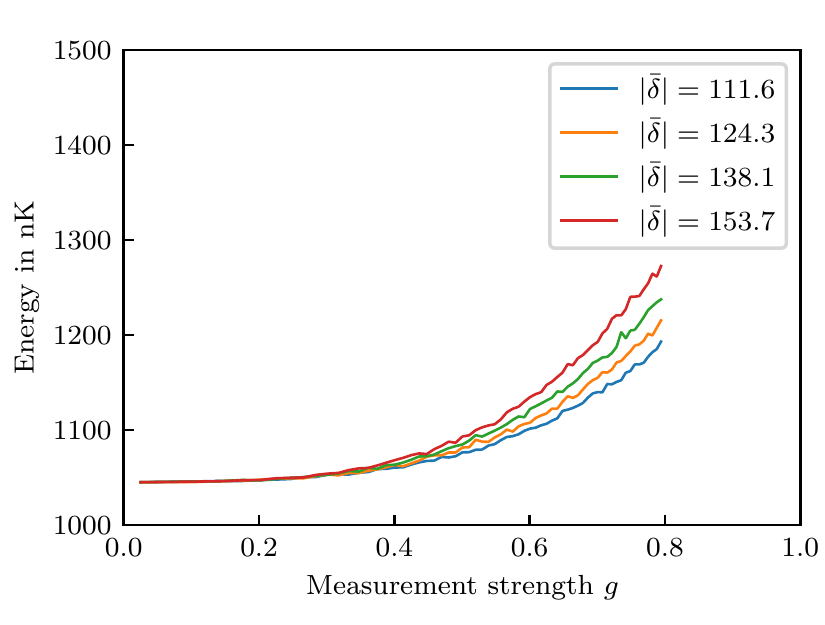}
\caption{
Modeled energy of a thermal cloud including spontaneous emission and an optical lattice for the parameters in Fig.~\ref{Fig4:Energy}(d).
The nominal lattice depth was selected to give the observed peak heating and was scaled to be proportional to $\bar I / \bar \delta$.
The peak lattice depths are $3.9 \Er$, $4.4\Er$, $4.8\Er$ and $5.4\Er$ for $\bar\delta = 111.6$, $124.3$, $138.1$ and $153.7$, respectively.
}
\label{FigSM:ScatModel}
\end{figure}

We modeled the added kinetic energy using a classical Monte Carlo simulation.
This simulation incorporates scattering as a stochastic process with atoms recoiling in the dipole emission pattern expected for the $\ket{F=2,m_F=2}\rightarrow\ket{F=3,m_F=3}$ cycling transition.
In addition, the simulation includes an optical lattice potential.

Our numerical approach solves the stochastic equations of motion using a first-order approximation for the derivatives~\footnote{Because higher order solvers require special care with stochastic equations, we opted instead for a first order solver with a small time step.}, with a time step selected to be much smaller than the scattering time.
This model should accurately describe the added energy for a weakly interacting thermal gas.
In addition we expect it to reasonably approximate the added kinetic energy for a BEC as the overall properties (such as the per-particle energy) of a BEC evolving in an optical lattice are well described using classical models~\cite{Huckans2009a}.

Figure~\ref{FigSM:ScatModel} plots the computed added energy associated with the thermal gas measurements in Fig.~\ref{Fig4:Energy}(d).
The lattice depth was selected to approximately account for the experimentally observed heating at $g=0.8$, and was scaled with $\bar I / \bar \delta$ away from this point (see caption).
These simulations show that even in principle these data do not collapse onto a single curve as a function of $g$.
The variation is comparable with the observed scatter in Fig.~\ref{Fig4:Energy}(d), suggesting that for the thermal gas (where lattice mitigation is ineffective) the remnant heating at red-detuning could be from stray lattice effects.

\bibliography{main}

\end{document}